\documentclass[sigconf]{acmart}
\usepackage{algorithm}
\usepackage{algorithmic}

\AtBeginDocument{%
  \providecommand\BibTeX{{%
    \normalfont B\kern-0.5em{\scshape i\kern-0.25em b}\kern-0.8em\TeX}}}

\copyrightyear{2021} 
\acmYear{2021} 
\setcopyright{acmcopyright}\acmConference[MM '21]{Proceedings of the 29th ACM
International Conference on Multimedia}{October 20--24, 2021}{Virtual Event, China}
\acmBooktitle{Proceedings of the 29th ACM International Conference on Multimedia
(MM '21), October 20--24, 2021, Virtual Event, China}
\acmPrice{15.00}
\acmDOI{10.1145/3474085.3475403}
\acmISBN{978-1-4503-8651-7/21/10}

\acmSubmissionID {mfp1374}

\begin{document}
\fancyhead{}

\title{SFE-Net: EEG-based Emotion Recognition with Symmetrical Spatial Feature Extraction}

\author{Xiangwen Deng}
\authornotemark[1]
\affiliation{%
  \institution{University of Electronic Science and Technology of China}}
  \email{sherwin_deng@std.uestc.edu.cn}

  \author{Junlin Zhu}
  \authornote{Equal contribution}
\affiliation{%
  \institution{University of Electronic Science and Technology of China}}
  \email{junlin_zhu@std.uestc.edu.cn}

  \author{Shangming Yang}
  \authornote{Corresponding author}
\affiliation{%
  \institution{University of Electronic Science and Technology of China}}
  \email{minn003@163.com}


\begin{abstract}
  Emotion recognition based on EEG (electroencephalography) has been widely used in human-computer interaction, distance education and health care. However, the conventional methods ignore the adjacent and symmetrical characteristics of EEG signals, which also contain salient information related to emotion. In this paper, a spatial folding ensemble network (SFE-Net) is presented for EEG feature extraction and emotion recognition. Firstly, for the undetected area between EEG electrodes, an improved Bicubic-EEG interpolation algorithm is developed for EEG channels information completion, which allows us to extract a wider range of adjacent space features. Then, motivated by the spatial symmetric mechanism of human brain, we fold the input EEG channels data with five different symmetrical strategies, which enable the proposed network to extract the information of space features of EEG signals more effectively. Finally, a 3DCNN-based spatial, temporal extraction, and a multi-voting strategy of ensemble learning are integrated to model a new neural network. With this network, the spatial features of different symmetric folding signals can be extracted simultaneously, which greatly improves the robustness and accuracy of emotion recognition. The experimental results on DEAP and SEED datasets show that the proposed algorithm has comparable performance in terms of recognition accuracy.
\end{abstract}



\maketitle


  \section{Introduction}
Emotion is a complicated physiological response of human beings, which plays an important role in our daily life and work\cite{article}. Positive emotions help us to improve human health and work efficiency, while negative emotions may cause health problems\cite{2018ABC}. Nowadays emotion recognition has been widely used in many scientific and technological fields, such as human-computer interaction, distance education, and health care, etc., and has gained wide attention of academic research.

In general, there are two different ways to recognize emotion. One is to directly consider non-physiological signals such as facial expressions\cite{2018Dominant}, speech\cite{8085174}, body gesture\cite{inproceedings}, text\cite{unknown} to construct models, which collect data in a non-invasive way. However, for this approach, it is difficult to obtain correct emotions if people deliberately mask their true feelings \cite{articleElicitation}. Another approach is to consider physiological signals such as heart rate\cite{s20030718}, skin conductivity\cite{s21031018}, respiration\cite{inbook}, and EEG to classify emotions. EEG, signaled by the central nervous system, is a direct response to brain activity and is more objective and reliable in capturing real human emotions.

EEG-based emotion recognition methods can be roughly partitioned into two parts: EEG feature extraction and emotion classification. Firstly, EEG features are extracted from the time domain, frequency domain, and time-frequency domain\cite{inproceedingsModel}. Then, they are used to train a classifier. Researchers mainly construct models using machine learning methods to deal with EEG emotion recognition tasks. However, the existing machine learning methods ignore the symmetric spatial characteristics of EEG signals, which may also contain salient information related to emotion. Deep learning has been widely applied in recent years, which has demonstrated its capabilities of temporal and spatial feature extraction in multi-channel EEG emotion recognition. 

For example, in \cite{articlefusion}, researchers applied GCN to extract graph domain
feature and LSTM to capture temporal features, and in \cite{articleJinpeng}, researchers also used convolutional neural network (CNN) to capture spatial features.

Although the existing emotion recognition models have achieved high accuracy, most models only consider spatial features of adjacent channels but ignore the symmetry of human brain in information processing. Neurological studies\cite{cui_2020} have shown that asymmetrical activities of the brain are effective in emotional processing, which indicates that there exist strong correlations for symmetrical channels in brain activities. Thus, how to capture adjacent and symmetrical features more effectively is still an interesting topic in the research of emotion recognition.

To apply the symmetric mechanism of human brain for EEG data processing, a spatial folding ensemble network named SFE-Net for the EEG feature extraction and emotion recognition is proposed in this paper. First, the input EEG channels data is folded with five different symmetrical folding strategies: the left-right folding, the right-left folding, the top-bottom folding, the bottom-top folding, and the entire double-sided brain folding. Additionally, to enable the SFE-Net to extract feature information of temporal and spatial features, an improved Bicubic-interpolation is proposed for EEG channel information completion, from which the proposed network can extract a wider range of adjacent and symmetrical spatial features of a human brain. Then 3D-CNN based spatial and temporal extraction and a multi-voting strategy of ensemble learning are employed to model a new neural network, from which the spatial features of different folds can be extracted simultaneously, which greatly improves the robustness and accuracy of emotion recognition.

In summary, the main contributions of this paper are as follows:

(1) An improved interpolation algorithm with five different symmetric folding strategies is proposed to reinforce the sample data, which is helpful for the model to extract more effective features of adjacent and symmetrical spatial features more effectively in the next step.

(2) A multi-voting strategy of 3D-CNN with ensemble learning is proposed for the folded signals, which greatly improves the robustness and accuracy of emotion recognition.

(3) Extensive experiments are conducted on two benchmark datasets. The experimental results show that the proposed algorithm has comparable performance in terms of recognition accuracy.

\section{Related Works}
The related literature to the proposed work can be organized into two categories.

\subsection{EEG Interpolation Algorithms}
Interpolation algorithms have been widely used in image processing for pixel filling. For example, Zheng et al.\cite{articleinterpolation} proposed an adaptive k-nearest neighbors search and random non-linear regression for image interpolation to improve picture resolution. Zhou et al. \cite{zhou_Quantum} implemented a bilinear interpolation algorithm on novel enhanced quantum image representation (NEQR). Khaledyan et al. \cite{2020Low} implemented a bilinear and bicubic image interpolation with low cost to efficiently improve the real-time image super-resolution.

Since EEG signals can also be viewed as images, many researchers have recently applied interpolation algorithms to enhance the spatial structure of EEG. For example, Huang et al. \cite{articleS-EEGNet} proposed a separable convolutional neural network(CNN) with bilinear interpolation in a brain-computer interface for EEG classification. Svantesson et al. \cite{articleVirtual} utilized CNN as an interpolation method to upsample and restore channels, which finally recreate EEG signals with higher accuracy. In the field of EEG-based emotion recognition, interpolation algorithms are still relatively scarce. Cho et al. \cite{Differenti} used RBF (Radial Basis Function) interpolation algorithm to construct a new 2D EEG signal frame, which combines with DE (Differential Entropy) characteristics to achieve higher accuracy. However, RBF is a global interpolation algorithm, which may result in large errors when the EEG value is in a local drastic variation. To overcome the limitation, a new interpolation approach called the improved Bicubic-EEG interpolation is proposed, which can avoid the issue that local drastic changes lead to precision decline.

\subsection{EEG Feature Extraction}

Neurological studies\cite{articleAC}, \cite{articleDiffering} have shown that emotional response is closely related to the cerebral cortex of the human brain, which is a 3D spatial structure and is not completely symmetrical between the left and right hemispheres. Consequently, emotion-related features of EEG signals, which in essence belong to time series, may not only exist on a temporal dimension but also exist on a spatial dimension for adjacent and symmetrical channels.
For temporal feature extraction, some researchers utilize deep learning models to obtain dynamic information based on raw EEG signals automatically. Soleymani et al. \cite{7112127} proposed an LSTM-RNN and continuous conditional random field algorithm to detect the emotional state of the subjects from their EEG signals and facial expressions. Anubhav et al. \cite{inproceedingsAnubhav} proposed an LSTM neural network that could efficiently learn the features from the band power of EEG signals. Fourati et al. \cite{Fourati} presented an Echo State Network (ESN), which applied a recursive layer to perform the feature extraction step directly from the raw EEG.

For spatial feature extraction, Yang et al. \cite{8489331} proposed a parallel convolutional recurrent neural network to extract spatial characteristics of EEG streams, which convert the chain-like EEG sequence into a 2D-like frame sequence. Li et al. \cite{articleJinpeng} presented a hierarchical convolutional neural network to capture spatial information among different channels based on 2D EEG maps. Song et al. \cite{8320798} proposed a dynamical graph convolutional neural network to explore the deeper-level spatial information for adjacent channels, which utilized a graph to model the multichannel EEG signals. 

Although these methods all reconstruct the EEG matrix based on the actual channel position of the brain to obtain spatial structure, they fill the undetected electrodes with zero in the matrix, which may destroy the original structure and lead to errors in classification. More importantly, they ignore the correlations for symmetrical channels, which also contain salient information in brain activities. To address the above-mentioned limitations, we propose the SFE-Net model, which takes into account adjacent and symmetrical spatial features simultaneously.

\section{Proposed Method}
By reconstructing the spatial characteristics of EEG channels with channel interpolation and spatial folding data preprocessing, 3D-CNN and ensemble learning are applied for feature extraction and data classification respectively.

\subsection{Reconstruction of the Spatial Characteristics of EEG Channels}
Using a one-dimensional vector to  represent each frame of EEG channels
may ignore the spatial structure. The two-dimensional EEG image for emotion recognition has been widely studied in recent years \cite{2018Continuous}. For example, the one-dimensional vector of each original EEG frame is 32*1 and the two-dimensional EEG image is 9*9. To improve the inconsistency and uncertainty of the spatial information between multiple adjacent channels, each frame of EEG channels is mapped to a matrix according to the structure of the human brain to obtain two-dimensional EEG images. As shown in Fig. 1, the left part 2 is a plan view of the international 10-20 system. And the blue-filled EEG electrodes are the test points implemented in the DEAP dataset. We relocate each time sampling to a two-dimensional electrode topology according to the spatial position of the electrode. The unused electrodes are ﬁlled with zero.

\begin{figure}[h]
\centering
\includegraphics[width = 0.95 \linewidth,height=4cm]{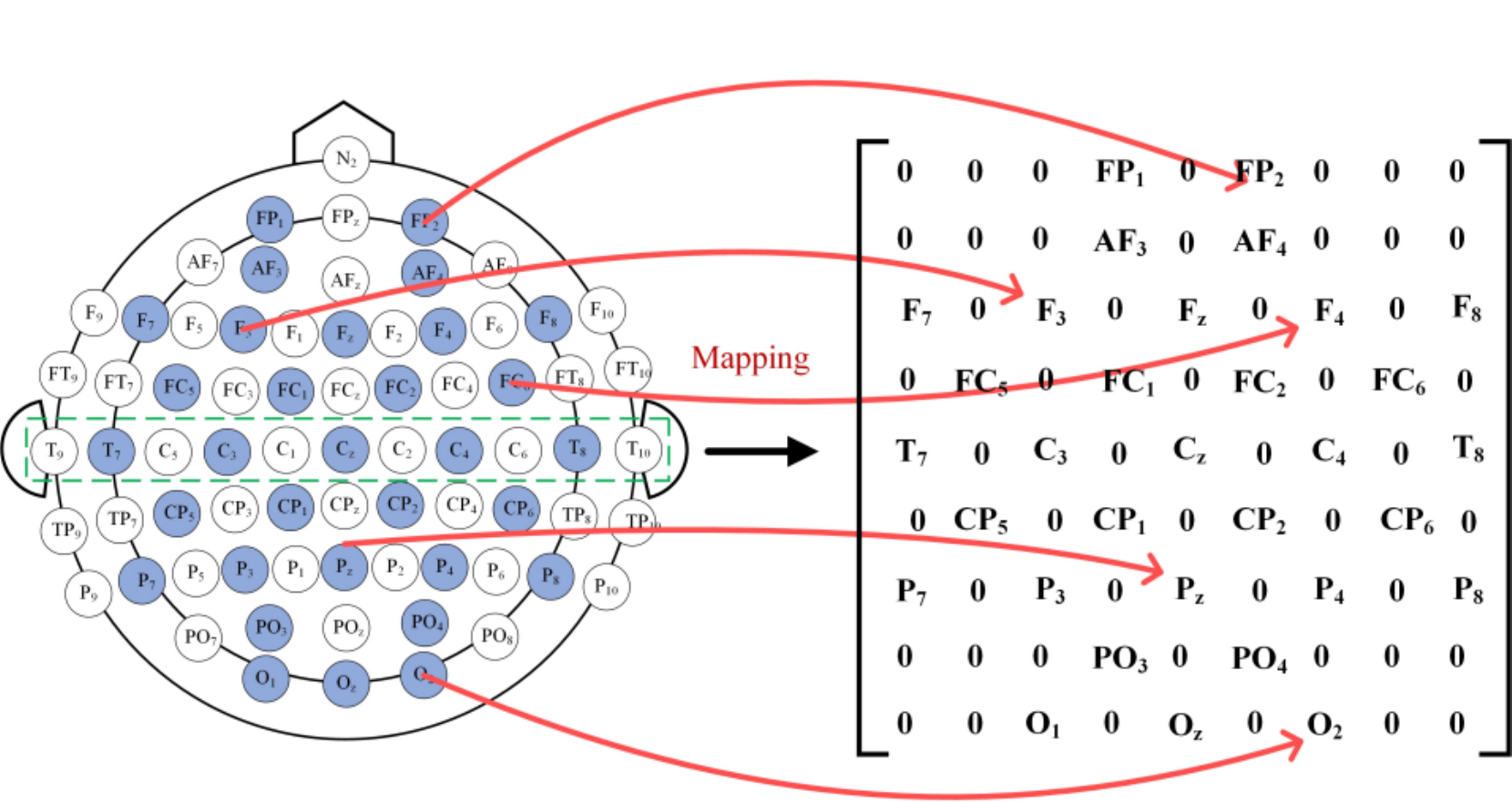}
\caption{The two-dimensional plane of the EEG electrode distribution map}
\end{figure}

In Fig. 1, $h$ and $w$ are the maximum number of electrodes which used vertically and horizontally. For the DEAP dataset and SEED dataset, $h$ = $w$ = 9. The experiment on the DEAP dataset has 15 electrodes, and the experiment on the SEED dataset has 62 electrodes. Then, channel filling and space folding based on a two-dimensional EEG will be performed for data completion.

\subsubsection{The Improved Bicubic-EEG Interpolation Algorithm}
\
\newline
\indent
Since there are certain unsampled parts between the brain electrical electrodes, we believe that this may ignore some critical information. So an improved interpolation algorithm is developed to reconstruct them. After comprehensively calculating the area around the blank space of the proposed algorithm, the new values can be obtained.

For the improved Bicubic-EEG interpolation algorithm, the inspiration came from the bicubic interpolation. In numerical analysis, bicubic interpolation is one of the most frequently used interpolation methods in a two-dimensional space. The classic algorithms of linear interpolation include: nearest neighbor method, bilinear interpolation, bicubic B-spline interpolation \cite{1993B} etc. Although the nearest neighbor method is simple and easy to implement, it will produce obvious jagged edges and mosaics in the new image. For bilinear interpolation, it has a smoothing function and can effectively overcome the drawbacks of the nearest neighbor method, but bilinear interpolation will degrade the high-frequency part of the image and make the image details blurred. As for the bicubic B-spline interpolation, compared to the bilinear interpolation, the transition and smoothness of the image will be focused more. However, the disadvantage is that the cost of computation is relatively high.

\begin{figure}[H]
\centering
\includegraphics[width = 0.95\linewidth,height=2.7cm]{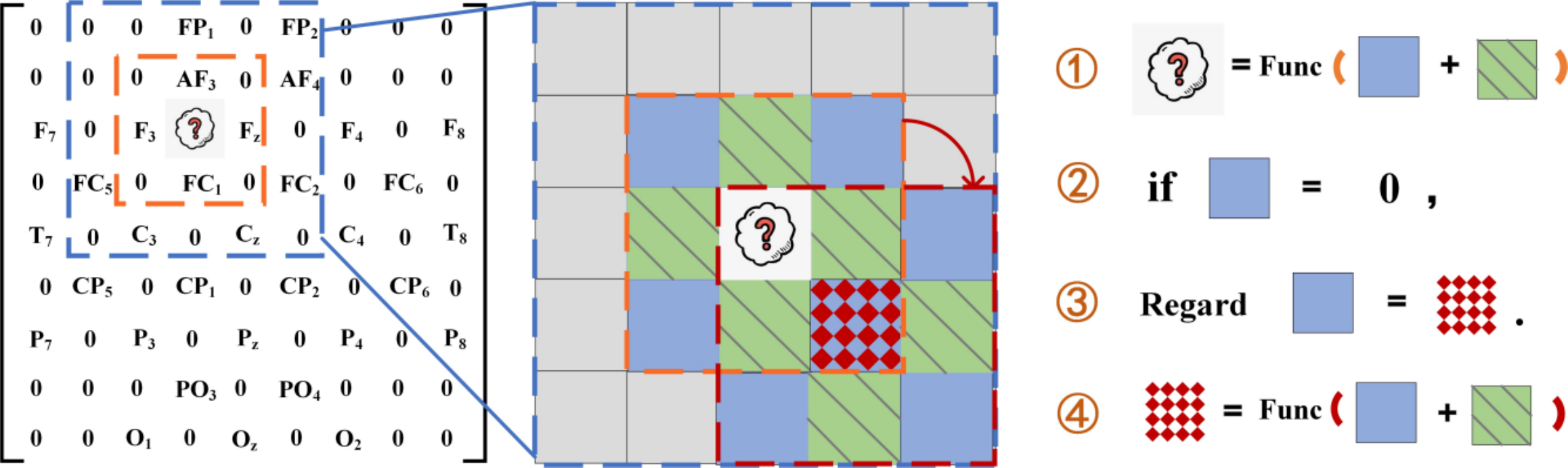}
\caption{The improved Bicubic-EEG interpolation scheme                                  
  }
\end{figure}

The original Bicubic-EEG interpolation method can increase the receptive field of adjacent areas to obtain higher efficiency of data completion. We re-optimized the algorithm according to the distribution of the EEG channels to solve the problem of the large cost of calculation in bicubic interpolation. Fig. 2 shows the improved Bicubic-EEG interpolation method. The right part in Fig. 2 is a simple profile of the algorithm.

Specifically, the detailed description of the improved Bicubic-EEG interpolation algorithm is shown as follows:

The original BiCubic function \cite{articleKeys} is

\begin{equation}
W(x) =
\begin{cases}
(a+2)\left\lvert x\right\rvert ^3 - (a+3)\left\lvert x\right\rvert ^2 +1       & \left\lvert x\right\rvert \leq 1 \\
a\left\lvert x\right\rvert ^3-5a\left\lvert x\right\rvert ^2+8a\left\lvert x\right\rvert -4a   & 1 < \left\lvert x\right\rvert  < 2 \\
0           & \left\lvert x\right\rvert  \geq 2 \text{,} \\  
\end{cases}
\end{equation}

\begin{figure*}[htb]
  \centering
  \includegraphics[width =0.8 \textwidth,height=7cm]{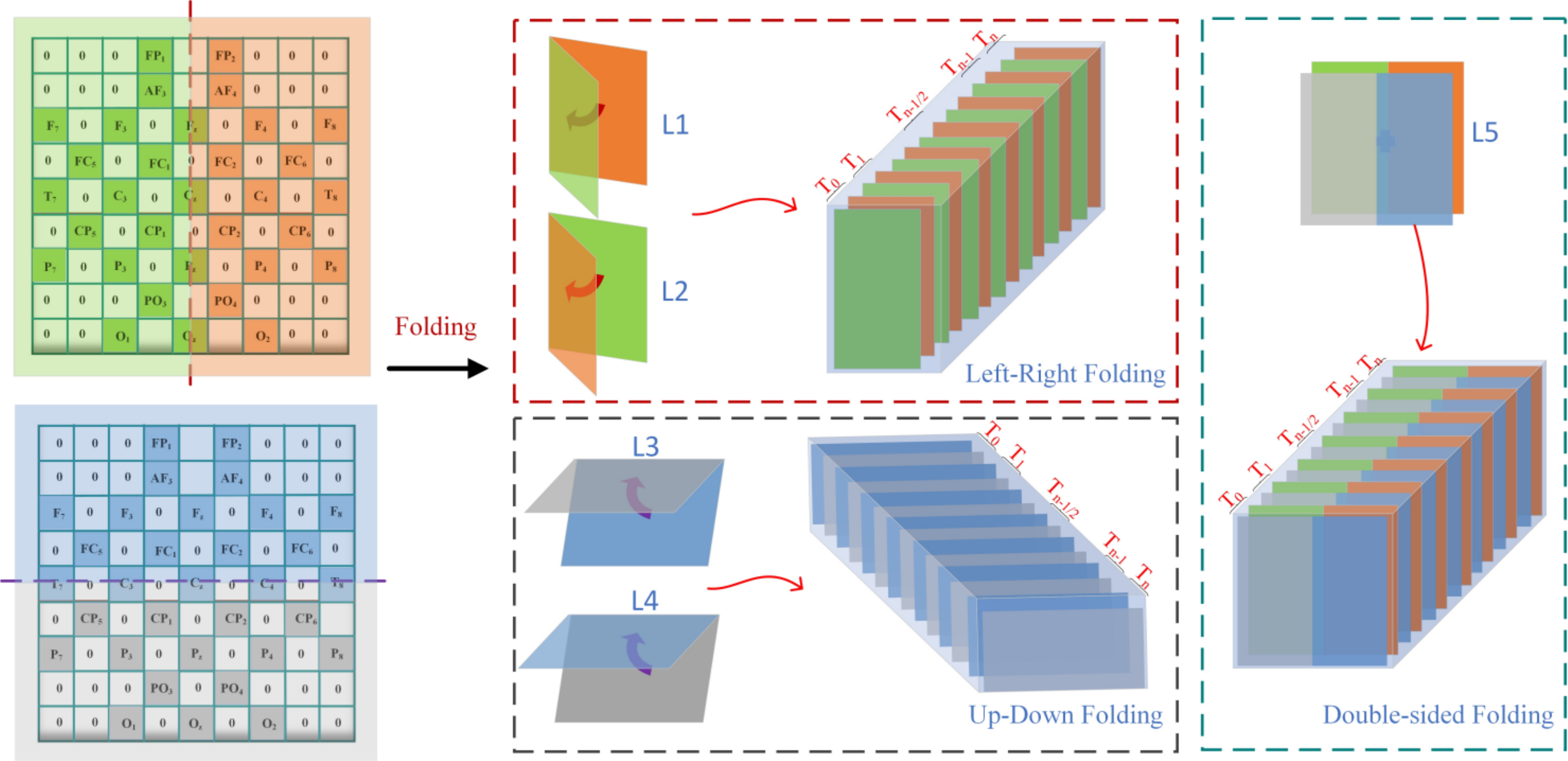}
  \caption{The proposed 5 folding approaches}
  \end{figure*}

  \noindent where $a$ is in $[-1,0]$. \cite{articleKeys} demonstrated that the best value of $a$ is -0.5. For the interpolated EEG channels $(x, y)$, the points $(x_i, y_j)$ is taken in a $3 \times 3$ neighborhood, $i,j = 0,1,2,\dots$ Then the interpolate points will be computed as follows,

\begin{equation}
  F(x, y) = \sum^2_{i=0}{\sum^2_{j=0}{f(x_i, y_j)W(x-x_i)W(y-y_j)}} \text{,}
  \end{equation}

  \noindent where $f(x, y)$ is the value to be filled at $(x, y)$, and $f(x_i, y_i)$ is the value of the EEG channels at $(x_i, y_i)$. Since $f(x_i, y_i)$ in Eq. (2) may have $f(x_i, y_i)=0$. In this case, the $f(x_i, y_i)$ is taken as the center, and again set a $3 \times 3$ neighborhood. After the neighborhood points are averaged, the value of $f(x_i, y_i)$ is obtained. In this case, $f(x, y)$ as defined as follows:

\begin{equation}
  F(x, y)=Avg\left(\sum^2_{i=0}{\sum^2_{j=0}{f(x_i, y_j)}}\right)  \text{,}
  \end{equation}

  \noindent where $Avg()$ is the average of the non-zero part of data.

  Thus, the interpolation can be obtained.

\subsubsection{Spatial Fold of EEG}
\
\newline
\indent
To better extract the feature of EEG channels, a data input model is constructed by fully considering the symmetrical information of various brain regions. All EEG channels are folded in half according to the distribution of the electrodes with different approaches. The folds include folding symmetrically from left to right or right to left, folding symmetrically from top to bottom or bottom to top, and folding the entire double-sided brain completely. In all, we have five types of folding. All the folding approaches are shown in Fig. 3. These five folding results are employed as the input data of the SFE-Net. With this input model, the ensemble learning in the next step can extract various symmetrical information at the same time.

Fig. 3 shows the process of the proposed fold model. Specifically, the folding method includes five categories: folding the left brain on the right brain, folding the right brain on the left brain, folding the forebrain on the hindbrain, folding the hindbrain on the forebrain, and folding the entire double-sided brain completely. 

For different folding strategies, the channel space organization is as follows: For left-right folds and right-left folds, the feature matrix size is $9 \times 5 \times 2$. For top-bottom folds and bottom-top folds, the feature matrix size is $5 \times 9 \times 2$. For the entire double-sided brain folding, the size of the feature matrix is $9\times9\times2$.

\subsection{The Proposed of SFE-Net}

The goal of the network is to fully extract the symmetrical features in the EEG to achieve accurate emotion recognition. The proposed network has three steps. First, the network folds the EEG data with five different approaches for input. Then the feature extraction capabilities of 3D-CNN is utilized to learn the temporal and spatial features of the adjacent and symmetric EEG channels of each folding mode.
Finally, the voting mechanism of ensemble learning is used to integrate these features into the classifier.

The framework of the SFE-Net model is presented in Fig. 4, from which we can see that SFE-Net is composed of three parts, including data folding, the 3D-CNN feature extracting, and ensemble learning.

\subsubsection{The 3D-CNN for Feature Extraction}
\
\newline
\indent
3D-CNN is a deep learning method, which is an extension of traditional CNN \cite{6165309}. It improves convolution and pooling operations, which makes it superior in constructing long sequence spatio-temporal feature models. 3D convolution generates a series of 3D feature volumes by processing 3D input, where the third dimension is modeled by continuous input time frames. 

Formally, the value at position $(x,y,z)$ on the $j$th feature map in the $i$th layer is given by

\begin{equation}
V_{ij}^{xyz}=f(\sum_m{\sum^{P_i-1}_{p=0}{\sum^{Q_i-1}_{q=0}{\sum^{R_i-1}_{r=0}{W_{ijm}^{pqr}V_{(i-1)m}^{(x+p)(y+q)(z+r)}}}}}+b_{ij}) \text{,}
\end{equation}

\noindent where $P_i$, $Q_i$ and $R_i$ are the sizes of the 3D kernel along the temporal dimension, $W_{ijm}^{pqr}$ is the $(p,q,r)$th value of the kernel connected to the $m$th feature map in the previous layer, and $f$ is the activation function of the network, which has a rectifying linear unit (Relu) to enhance network performance.

Due to the dependence between long-time sequence fragments such as voice, video, and EEG signals, studies show that the utilization of 3D-CNN can improve the accuracy and robustness of the model. In addition, three-dimensional convolution operations can visualize and model the spatial correlation between video frames or pixels. For example, 3D-CNN is utilized for action recognition in \cite{2015Learning}. Since the EEG channels signal also has both time and space features, 3D-CNN is applied to fully extract time and space features from the 3D EEG input cube we constructed. 

\subsubsection{Ensemble Learning and Voting Mechanism}
\
\newline
\indent
The ensemble learning implements independent training based on several learning algorithms. Then, it flexibly combines the basic learners of the learning results.

\begin{figure*}[h]
  \centering
  \includegraphics[width =0.9 \textwidth,height=7.5cm]{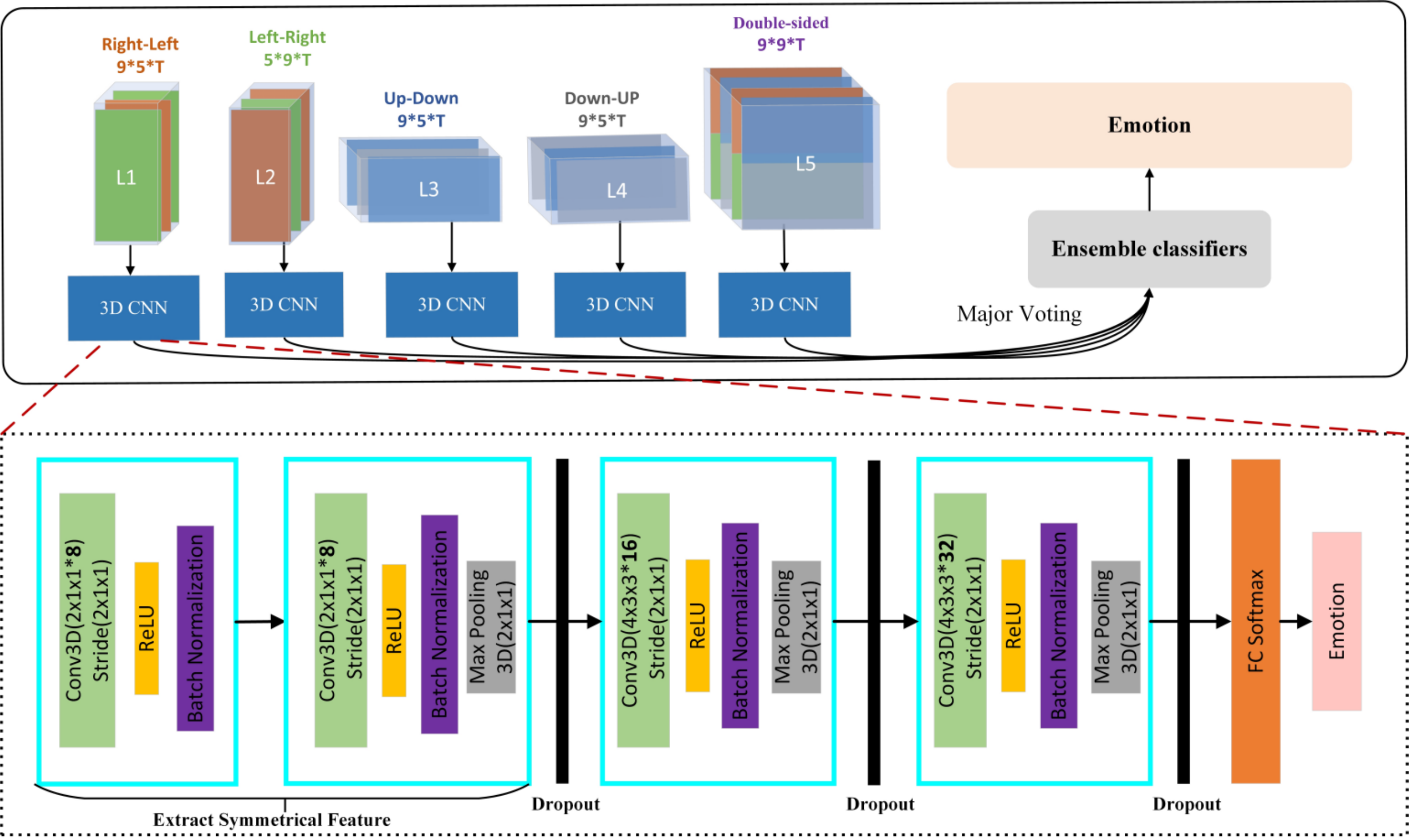}
  \caption{The structure of SFE-Net}
  \end{figure*}

Since the learning results for different algorithms generally have certain differences, the combined learning result is usually better than that of an individual algorithms. The design of each basic classifier in ensemble learning should be different from each other. Generally, a basic classifier can be constructed by using different input features, training sets, and learning algorithms \cite{Padilha2016A}. To have some difference for different algorithms \cite{10.1007/3-540-45014-9_1}, we consider that not only the left and right sides of the brain may have symmetrical regional features, but the front and rear sides of the brain may also have symmetrical regional features. Thus ensemble learning can be applied to synthesize the characteristics of front and rear symmetry and the left and right symmetry at the same time. In the proposed model, ensemble learning is employed to integrate five different EEG folding approaches. Thus, ensemble learning can extract different channels of EEG symmetry features simultaneously. Therefore, this approach that integrates all the symmetric features helps to improve the robustness of the network.

The proposed SFE-Net framework is shown in Fig. 4. Five CNN classifiers are employed, including CNN1 \dots CNN5. Every CNN classifier is a basic learning algorithms. From all the five CNN, a powerful algorithms is formed through a diversified voting strategy. Each learning algorithm $y_{i} $ will predict the result $Y(x)$ according to the category set {$C_{1}$, $C_{2}$\dots, $C_{N}$}. The integration strategy is formulated as follows:

\begin{equation}
Y(x)=\mathop{\arg\max}_{j}{\sum_{i=1}^{T} {y_{i}^{j}{(x)}}} \text{,}
\end{equation}

\noindent where the N-dimensional $\overrightarrow{vector} \left(y_{1}^{j}{(x)}, y_{2}^{j}{(x)},...y_{N}^{j}{(x)}\right) $ represents the predicted output of $y_{i}$ on the sample $x$, $y_{i}^{j}{(x)}$ represents the output of $y_{i}$ on the category $C_{j}$, $T$ is the number of learners, $j$ is the number of categories, and its value is 2 or 4.

\section{Experiments}

\subsection{Materials}

In the experiment, the DEAP dataset \cite{2012DEAP} and the SEED dataset \cite{2019Identifying} are selected to evaluate the performance of the network. These two datasets are widely used in the field of emotion recognition.

DEAP: The DEAP dataset is a multi-modal dataset developed by Koelstra and colleagues, including 32-channel EEG signals and 8-channel peripheral physiological signals. Researchers recorded the signals from 32 healthy volunteers when they watched 40 one-minute emotional music video clips. The researchers conducted 40 trials on each volunteer and asked them to rate the Arousal, Valence, Dominance and Liking of each video from 1 to 9. Each trial contains 63s signals and the rest 3s are for the baseline signals. The EEG signal is sampled at 512hz and then down-sampled to 128hz. The structure of the DEAP dataset is shown in Table 1.

\begin{table}[htb]
  \centering
  \caption{The structure of the DEAP dataset}
  \begin{tabular}{ccc}
  \toprule
  Array name & Array shape & Array contents \\
  \midrule
  Data & $40\times40\times8064$ & $Videos\times Channels\times Clips$\\
  Labels & $40\times4$ & $Videos\times Labels$ \\
  \bottomrule
  \end{tabular}
  \end{table}

  \begin{table}[htb]
    \centering
    \caption{The structure of the SEED dataset}
    \begin{tabular}{ccc}
    \toprule
    Array name & Array shape & Array contents \\
    \midrule
    Data & $45\times 62\times 12000$ & $Videos\times Channels\times Clips$\\
    Labels & $45\times 1$ & $ Videos\times Labels$ \\
    \bottomrule
    \end{tabular}
    \end{table}
  
\begin{table*}[htb]
  \centering
  \caption{\centering{The comparative experiment results}}
  \begin{tabular}{cccccc}
  \hline
  Method                         & Model             & \textit{DEAP Arousal(\%)} & \textit{DEAP Valence(\%)} & \textit{SEED Three-category classification (\%)} & Year          \\ \hline
  \cite{2017MultimodalTang}    & Bimodal-LSTM      & /                         & 83.53                     & 93.97                                            & 2017          \\
  \cite{2018Correlated}        & CAN               & 85.62                     & /                         & 94.03                                            & 2018          \\
  \cite{2019Multi-method}      & ST-SBSSVM         & /                         & 72.00                     & 89.00                                            & 2019          \\
  \cite{2018Continuous}        & 3D-CNN+DE          & 89.45                     & 90.24                     & /                                                & 2019          \\
  \cite{Haiping2019Multimodal} & ECNN & /                         & 82.92                     & /                                                & 2019          \\
  \cite{2020Autoencoder}       & SAE-DNN               & 89.49                     & 92.86                     & 96.77                                            & 2020          \\
  \cite{2021Multi-Scale}       & Ensemble Learning &/                           & /                         & 82.75                                            & 2020          \\
  \cite{2020EFDMs}             & CNN+EFDMs         & 82.84                     & /                         & 90.59                                            & 2020          \\
  \cite{2020Cross-Subject}     & CNN               & 72.81                     & /                         & 78.34                                            & 2020          \\
  \cite{2020EEGSpiking}        & SNN               & 74.00                     & 78.00                     & 96.67                                            & 2020          \\
  \cite{2020Latent}            & Autoencoder       & 79.89                     & 76.23                     & 85.81                                            & 2020          \\
  \textbf{Ours}                  & \textbf{SFE-Net}  & \textbf{91.94}            & \textbf{92.49}            & \textbf{99.19}                                   & \textbf{2021} \\ \hline
  \end{tabular}
  \end{table*}

SEED: It is a collection of EEG records prepared by the Brain-Inspired Computing and Machine Intelligence (BCMI) Laboratory of Shanghai Jiaotong University. A total of 15 segments are selected to stimulate (neutral, negative and positive) emotions. Each stimulus includes a 5-second movie prompt, 4-minute editing, 45-second self-assessment, and a 15-second rest time. A total of 15 Chinese subjects participated in this study. Each subject needs to perform the same experiment three times on different dates to ensure the authenticity and universality of the extracted EEG data. A total of 45 EEG data were recorded. Tags were assigned based on the content of the clip (-1 for negative, 0 for neutral, and 1 for positive). The data was collected through 62 channels. These channels were placed according to a 10-20 system and downsampled to 200 Hz. Since the duration of each video is different, to be unified, the middle 60s of each video was intercepted as the duration. The structure of the SEED dataset is shown in Table 2.

\subsection{Implementation Details}

The structure of SFE-Net is shown in Fig. 4. The proposed SFE-Net model is mainly composed of 2 parts. Firstly, the model with the space-time extraction capabilities of 3D-CNN was trained parallelly. Secondly, we integrated all the different folding results and obtained the final result of the voting decision.

The SFE-Net model includes four convolutional layers. Since the time dimension is doubled after folding, the purpose of the first layer of convolution is to extract features in the time dimension and reduce the number of parameters. To reduce the loss of information of the input data, zero padding is used in each convolutional layer. After each convolution, batch-normalization is adopted. Then RELU activation function is utilized to increase nonlinearity. The pooling size of the largest pooling layer is $2\times 1\times 1$, which can reduce the size of the data in the time dimension. The feature maps of each layer are 8, 8, 16, and 32 respectively. Finally, a fully connected layer is designed to map the 32 feature maps into a feature vector of 512 dimensions. After regularization, set dropout = 0.5 to avoid over-fitting in a fully connected layer. In each convolutional layer, L2 regularization may prevent over-fitting, and L2=0.001. The Adam optimizer can minimize the cross-entropy loss function. The learning rate is set to 0.001. The learning epoch is 100. 
For the subject-dependent classification of 32 subjects, the dataset is split in patient-wise and for each subject, 80\% of EEG data are randomly used for training and 20\% for testing. For the subject-independent classification of 32 subjects, the data of the 32 subjects are integrated and randomly selected 80\% for training and 20\% for testing. 5 fold cross-validation methods is adopted in the experiments, in which the dataset is split into 5 parts and we circularly select 4 parts for training and the rest for testing.

\section{Results and Discussion}
\subsection{Results for Subject-independent Classification}
To show the performance of the proposed method, the proposed algorithm will be compared with the latest methods on the DEAP and SEED datasets, including CNN-LSTM, CNN+EFDMs, CAN, and some other methods. The special channel setting is used to extract the hidden features of symmetrical channels.

Comparing with multimodal methods such as ECNN, Bimodal-LSTM, the recognition accuracy of the proposed model increased by more than 4\%, which indicates that the proposed method is not just an efficient single-mode, but has better performance.

Comparing with manual feature extraction methods such as SNN, ST-SBSSVM, and 3D-CNN+DE, our method can construct feature data automatically on raw data.

These methods either have a spatial setting to extract spatio-temporal information, or manually construct special feature data for identification. However, the symmetrical channel information of EEG signals is ignored. SFE-Net also benefits from the ensemble learning of different folding approaches and comprehensive voting decision-making strategies. Extracting more symmetrical information at the same time is equivalent to an efficient data enhancement. Voting decision-making improves the robustness of the model. Table 3 shows the comparison results of different methods on the DEAP dataset and SEED dataset. From this table, we can see that SFE-Net can achieve very competitive experimental results. The accuracy on DEAP has reached 91.94\% in Arousal and 92.49\% in Valence. It also reached 99.19\% on the SEED dataset. Compared with existing EEG recognition methods, the accuracy has significantly improved by SFE-Net.

\subsection{Results for Subject-dependent Classification}

In order to further verify the effectiveness of SFE-Net approach, we compare the proposed 
method with the state-of-the-art methods, including decision tree (DT) \cite{2018Continuous},dynamic graph convolutional neural 
network (DGCNN) \cite{2018EEGDynamical}, multi-grain cascade forest (gcForest) \cite{9096541}, continuous 
CNN (Cont-CNN) \cite{2018Continuous}, 
Support Vector Machine (SVM) \cite{Bernhard2008Least}, 
Multilayer Perceptron (MLP) \cite{2018Continuous}, 
and CNN-RNN \cite{8489331}.

DGCNN was proposed by Song et al \cite{2018EEGDynamical}, which can dynamically learn the internal relationship between different EEG channels represented by the adjacency matrix, thereby classifying EEG emotions. Cont-CNN is a convolutional neural network with no pooling operation, which takes a constructed 3D EEG cube as input. The 3D EEG cube is a three-dimensional representation that combines the DE features with the spatial information between the electrodes \cite{2018Continuous}. gcForest is a model based on deep forest, Cheng et al. \cite{9096541} applied it to emotion recognition based on EEG. Yang et al. also constructed features based on 2D EEG frames \cite{8489331}. CNN-RNN is a hybrid neural network. It extracts spatial and temporal features from the constructed 2DEEG frame and 1D EEG sequence, respectively. To be fair, we will directly take the results in their literature as a comparison \cite{2020Multi-Capsule}.

From the results, first of all, the proposed method is far superior to the recognition effect of basic neural networks such as MLP and SVM. Then, compared with the two latest CNN-based methods (Cont-CNN and CNN-RNN), the proposed method also shows obvious advantages. The results on the DEAP dataset are shown in Fig. 5, Fig. 6. and the results on the SEED dataset are shown in Fig. 7.

\begin{figure}[htb]
  \centering
  \includegraphics[width = \linewidth,height=6cm]{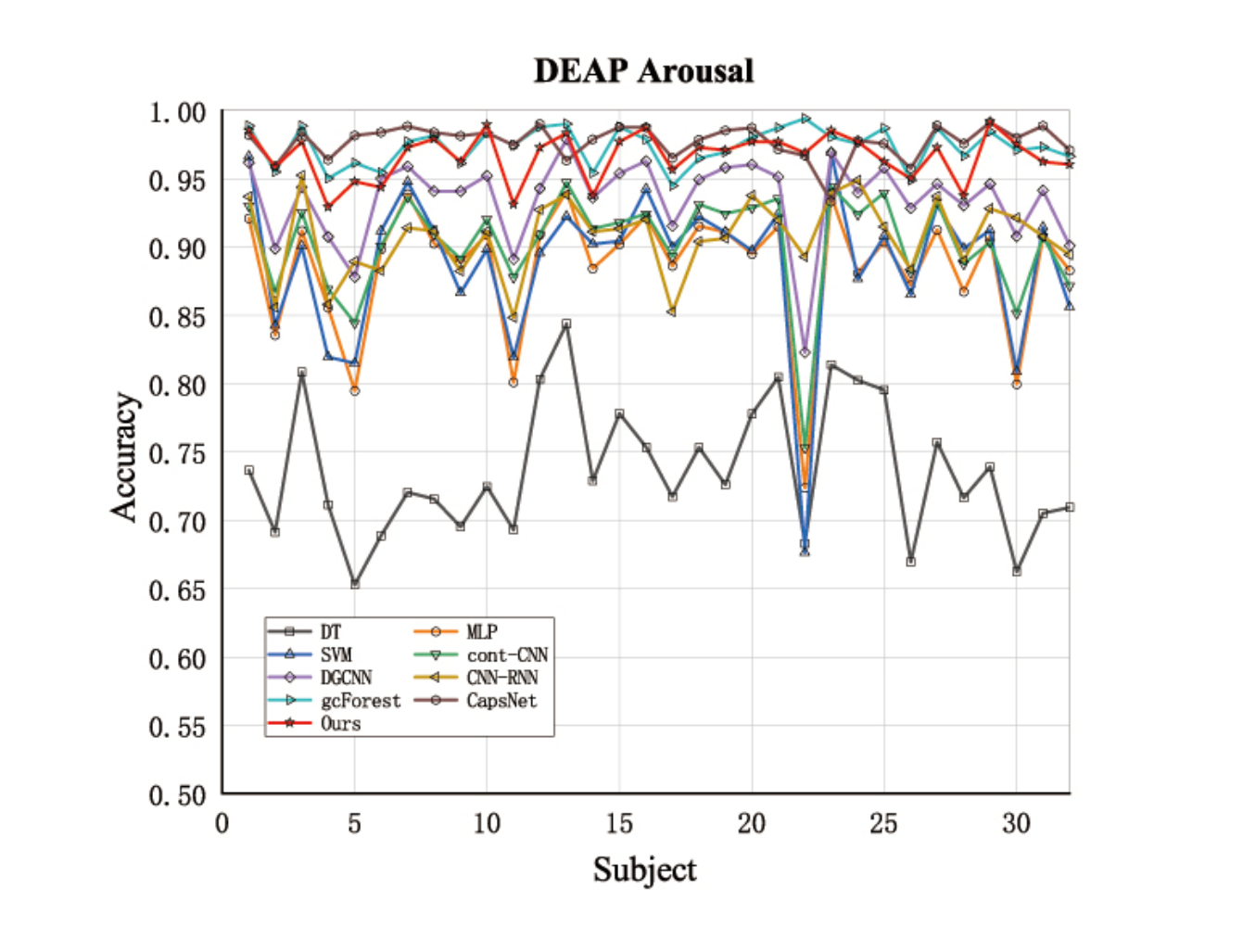}
  \caption{Arousal classification accuracies (\%) of each subject for different methods on the DEAP dataset}
  \end{figure}
  
  \begin{figure}[htb]
  \centering
  \includegraphics[width =  \linewidth,height=6cm]{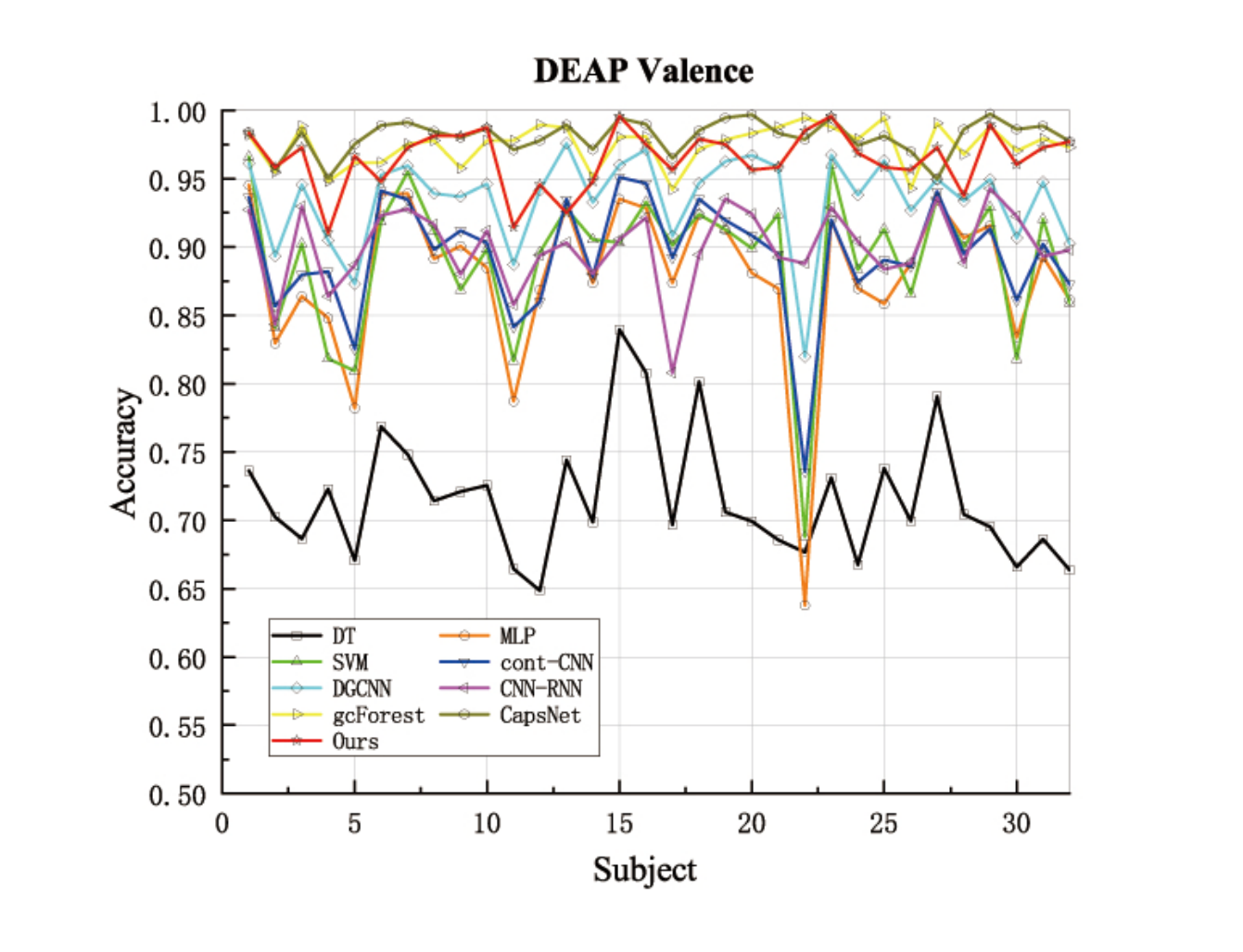}
  \caption{ Valence classification accuracies (\%) of each subject for different methods on the DEAP dataset}
  \end{figure}

  \begin{figure}[htb]
    \centering
    \includegraphics[width =\linewidth,height=6cm]{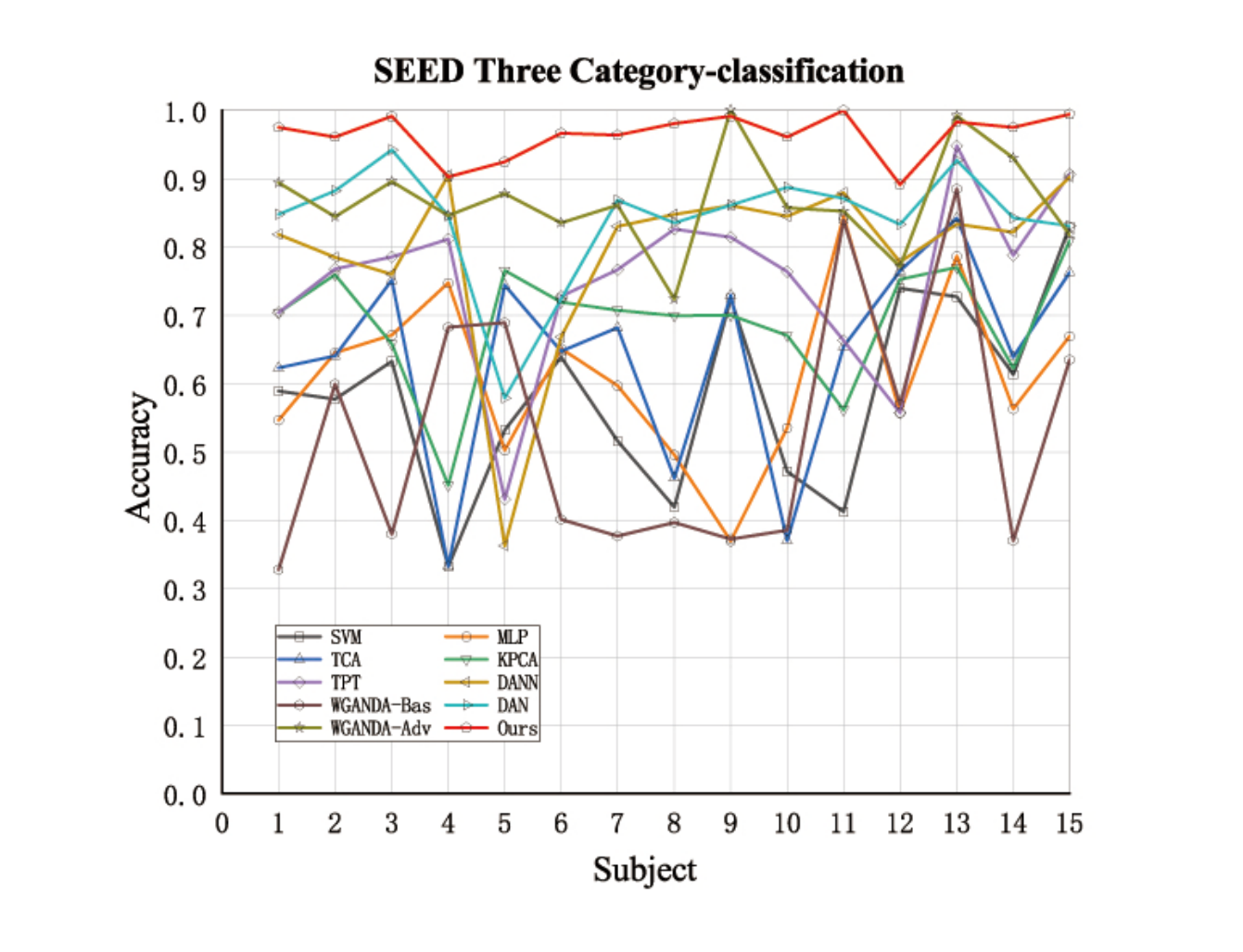}
    \caption{Three-category classification accuracies (\%) of each subject for different methods on the SEED dataset
    }
    \end{figure}

\subsection{Discussion}

The proposed model can achieve high accuracy, mainly because it developed ensemble learning with the improved Bicubic-EEG interpolation algorithm and EEG
folding. This network can effectively learn higher-level adjacent and symmetrical spatial features related to emotions. In this section, three types of comparative experiments are conducted to evaluate the performance of the proposed emotion recognition method. At first, a comparison of five different folding strategies is conducted to explore the best folding approach and evaluate the effectiveness of the method. Then, interpolation algorithms are employed to compare the influence of different algorithms on the experimental results to verify the effectiveness of the improved Bicubic-EEG interpolation algorithm. Finally, ensemble learning is used as a comparison to further explore the best ensemble strategy by integrating different folding methods. 

\subsubsection{Various Folding Methods}
\
\newline
\indent
To show the advantages of spatial folding in this study, we have five types of folded original EEG data and unfolded original EEG data for comparison. Thus, the sample data
for emotion recognition after spatial folding can be obtained.
The other experimental settings and network parameters remain unchanged.

\begin{table}[htb]
\centering
\caption{\centering{The different ways to fold signals on the DEAP dataset}}
\begin{tabular}{cccc}
\hline
\textit{Sub} & \textit{DEAP Arousal} & \textit{DEAP Valence} & \textit{SEED} \\ \hline
No\_fold     & $84.49\pm1.22$     & $90.12\pm1.71$      & $78.73\pm1.37$                        \\
Left\_fold   & $84.98\pm1.24$     & $87.75\pm1.59$      & $96.42\pm0.92$                        \\
Right\_fold  & $85.61\pm1.95$     & $87.74\pm1.57$      & $98.33\pm1.19$                        \\
Up\_fold     & $86.56\pm0.91$     & $87.50\pm0.82$      & $97.79\pm0.99$                        \\
Down\_fold   & $85.75\pm2.03$     & $87.37\pm1.08$      & $97.69\pm1.28$                        \\
Full\_fold   & $85.62\pm1.08$     & $87.80\pm1.79$      & $97.87\pm1.02$                        \\ \hline
\end{tabular}
\end{table}

The results are shown in Table 4. We can see that the difference is not very high on DEAP Valence. The average increase on DEAP Arousal is 1.21\%, and the maximum increase is 2.07\%. However, the folding methods have achieved amazing results on the SEED dataset, with an average increase of 18.89\%, and a maximum increase of 19.6\%. The results show that the method using half-folding to reconstruct the spatial features of the original data can greatly improve the ability of 3D-CNN to extract symmetric features.

\subsubsection{Channel Interpolation Algorithm}
\
\newline
\indent
To show the effectiveness of the developed Bicubic-EEG interpolation algorithm, the improved Bicubic-EEG interpolation algorithm is compared with the Bilinear interpolation algorithm and unfilled raw EEG data. The experimental results are shown in Fig. 8 and Fig. 9.

\begin{figure}[htb]
  \centering
  \includegraphics[width = 0.8 \linewidth,height=4.5cm]{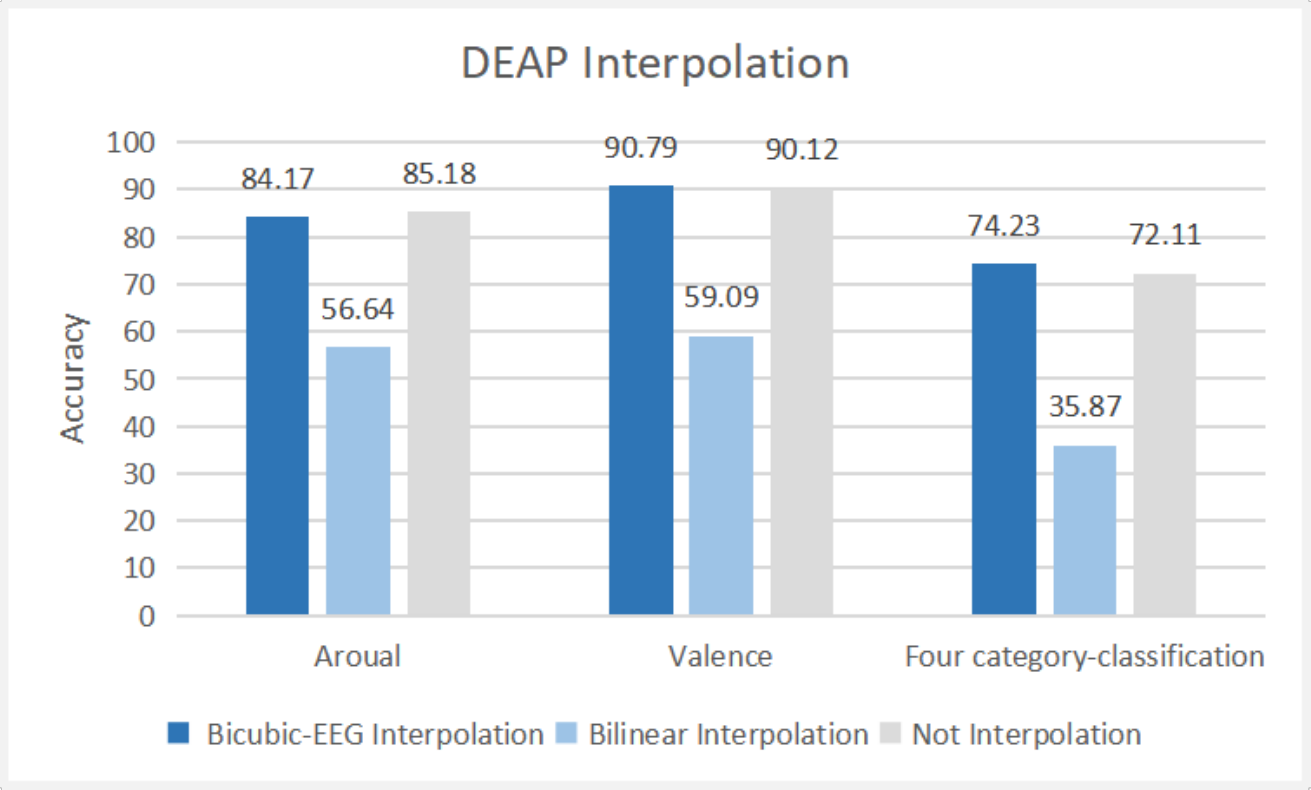}
  \caption{Performance comparison between different interpolation methods on the DEAP dataset}
  \end{figure}
  
  \begin{figure}[htb]
  \centering
  \includegraphics[width = 0.8 \linewidth,height=4.5cm]{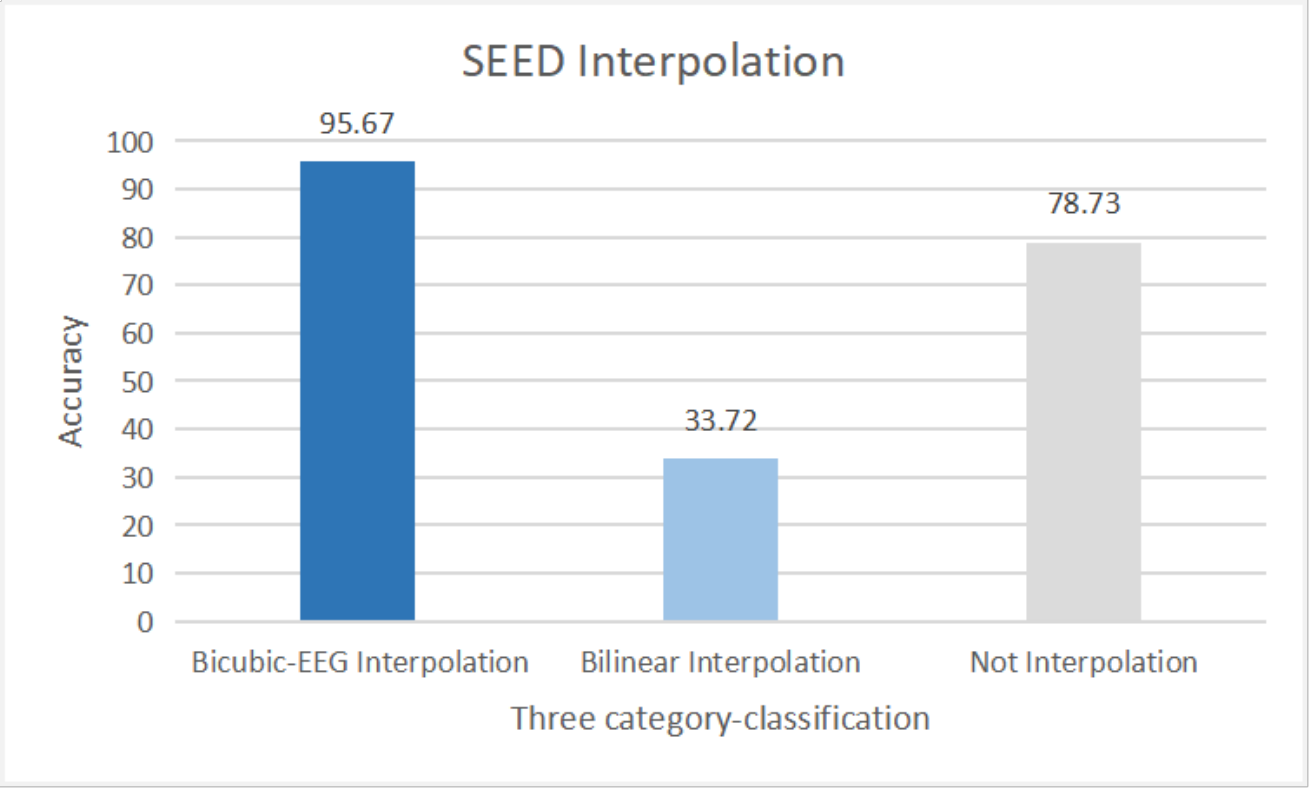}
  \caption{Performance comparison between different interpolation methods on the SEED dataset}
  \end{figure}
  
From the comparative experiments, we can see that the improved Bicubic-EEG interpolation algorithm has achieved a 2.12\% improvement on DEAP four-category classification (low/high arousal and low/high valence). However, amazing progress on the SEED dataset is achieved, which increases the accuracy by 16.94\%. At the same time, all experiments have demonstrated that the improved Bicubic-EEG interpolation algorithm is better than the general linear interpolation algorithm. The results show that the proposed method improves 3D-CNN's ability to extract adjacent spatial features in unsampled regions of the EEG data.

\subsubsection{Different Ensemble Strategies}
\
\newline
\indent
To show the efficiency of integrated learning strategy in ensemble learning, two types of comparative experiments are conducted. In the first one, voting is not applied but the average strategy is employed to integrate different classifiers. The second one is to repeat the same folding method five times for voting integration, instead of the previous voting integration with five different foldings.

\begin{table}[htb]
\centering
\caption{Test results of different folding strategies under ensemble learning}
\begin{tabular}{cccc}
\hline
DEAP                   & \textit{Arousal(\%)} & \textit{Valence(\%)} & \textit{Four\_class} \\ \hline
Nofold\_vote           & 88.43                & 89.38                & 82.61                \\
Leftfold\_vote         & 90.50                & 91.33                & 84.18                \\
Upfold\_vote           & 89.60                & 90.33                & 82.46                \\
Ourfold\_average       & 86.24                & 88.07                & 77.85                \\
\textbf{Ourfold\_vote} & \textbf{91.94}       & \textbf{92.49}       & \textbf{86.30}       \\ \hline
\end{tabular}
\end{table}

The experimental results are shown in Table 5, from which we can see that for the three evaluation indicators DEAP Arousal, DEAP Valence, and DEAP four-category classification, the the proposed model has 5.7\%, 4.42\%, and 8.45\% more than that of the average strategy obtained respectively. All of them are higher than the average strategy. For these three evaluation indicators, on average, the voting results of the proposed five different folding methods are 2.43\%, 2.14\%, and 3.22\% higher than that of the same folding method five times. From the results, we can see that the voting strategy obviously performs better than the average strategy. And the  integration of using five different folding methods performs better than the integration of using the same folding method five times repeatedly. These results show the effectiveness of the proposed ensemble method.

\section{Conclusion}
A spatial folding ensemble network (SFE-Net) is presented for electroencephalography feature extraction and emotion recognition. By combining time, regional symmetry, and regional adjacent information, this model can automatically extract features to achieve emotion recognition.

The average accuracies of the proposed model for the Valence classification and Arousal classification on the DEAP dataset are 91.94\%, 92.49\%, respectively. Specifically, the accuracy is 99.19\% on the SEED dataset. Thus, the results show that the algorithm we poposed can obtain state of art performance, which significantly demonstrate the effectiveness of the algorithm in learning spatial symmetry and spatial region information. However, in the proposed work, only single-mode experiments were performed. In the future, we will explore how to construct a multi-modal model to better extract various features.

\section{Acknowledgments}

This work was supported in part by the National Natural Science Foundation of China (NSFC) under Grant 61572107, in part by the National Key Research and Development Program of China under Grant 2017YFC1703905, 2018AAA0100202, in part by the National Science and Technology Major Project of the Ministry of Science and Technology of China under Grant 2018ZX10715003.


\newpage
{\small
\bibliographystyle{ACM-Reference-Format}
\bibliography{sample-sigconf.bib}
}

\appendix
\end{document}